\documentclass[10pt,english,reqno]{bourbaki}
\usepackage{amsmath}
\usepackage{amsthm}
\usepackage{amsfonts}
\usepackage{amssymb}
\usepackage{epic}
\usepackage{eepic}
\usepackage{times}

\theoremstyle{plain}

\theoremstyle{remark}
\newtheorem{rem}{Remark}

\newcommand{\sst}{\scriptscriptstyle}

\newcommand{\beq}{\begin{equation}}
\newcommand{\eeq}{\end{equation}}

\newcommand{\pa}{\partial}

\newcommand{\ra}{\rightarrow}

\newcommand{\ti}{\times}

\newcommand{\fr}[2]{{\textstyle \frac{#1}{#2} }}

\newcommand{\fsl}{{\mathfrak s}{\mathfrak l}}

\newcommand{\bra}{\langle}
\newcommand{\ket}{\rangle}
\newcommand{\al}{\alpha}

\newcommand{\ga}{\gamma}
\newcommand{\Ga}{\Gamma}
\newcommand{\de}{\delta}
\newcommand{\De}{\Delta}

\newcommand{\up}{\Upsilon}

\newcommand{\bx}{\bar{x}}

\newcommand{\bz}{\bar{z}}

\newcommand{\CD}{{\mathcal D}}

\newcommand{\CF}{{\mathcal F}}
\newcommand{\CG}{{\mathcal G}}

\newcommand{\CU}{{\mathcal U}}

\newcommand{\BR}{{\mathbb R}}

\newcommand{\BC}{{\mathbb C}}
\newcommand{\BP}{{\mathbb P}}

\newcommand{\smal}{\begin{smallmatrix} \al_3  \\
\al_4  \end{smallmatrix}}

\newcommand{\rf}[1]{(\ref{#1})}

\newcommand{\aufz}
{\begin{list}{$\bullet$}{\topsep0cm \itemsep0cm \parsep0cm}}
\newcommand{\eaufz}{\end{list}}
\newcounter{num}
\newcommand{\remlst}{\begin{list}
{(\arabic{num})}{\usecounter{num}\topsep0cm \itemsep0cm \parsep0cm}}
\begin{document}
\thispagestyle{empty}
\date{August 2001}
\title{Crossing symmetry in the $H_3^+$ WZNW model}
\author{J. Teschner}
\address{Inst. f\"ur theoretische Physik,\\
Freie Universit\"at Berlin,\\
Arnimallee 14,\\
14195 Berlin,\\
Germany}

\maketitle
\section{Introduction}

The $H_3^+$ WZNW model plays an important role in the study of 
string theory on $AdS_3$ (see e.g. \cite{MO}\cite{KS}\cite{GK} 
and references therein). It also has been proposed to be relevant for
the study of the integer quantum Hall effect \cite{Zi}\cite{QH}. One may 
furthermore view it as a prototypical example for a conformal field
theory with continuous spectrum of primary fields (noncompact CFT). 
 
The study of the $H_3^+$ WZNW model was initiated by Gawedzki,
who determined the spectrum of the theory from a path-integral 
calculation of the torus partition function \cite{Ga}.
The three point function for the $H_3^+$ WZNW model was calculated in 
\cite{T1}\cite{T2} by assuming decoupling of singular vectors in 
degenerate representations of the current algebra. 
Three point function and current algebra symmetry would fully
characterize the whole theory (see \cite{T3} for a discussion of the
corresponding issue in the case of Liouville theory).
However, so far there was no proof that the three point function 
from \cite{T1}\cite{T2} actually leads to a consistent theory.
Consistency of the theory boils down to proving crossing symmetry
of the four-point function that, as shown in \cite{T2}, can be uniquely
constructed out of the three point functions by means of the current algebra
symmetry.

The aim of the present paper is to show that crossing symmetry
of the four point function in the $H_3^+$ WZNW model follows
from similar properties of a five point function in Liouville theory.
A proof of 
the fact that locality and crossing symmetry hold in Liouville theory
was outlined in \cite{T3} building upon \cite{PT}.
One may therefore regard the results of \cite{T1}\cite{T2} as providing 
a solid construction of the $H_3^+$ WZNW model in genus zero, the 
consistency of which is established in the present note on the basis
of results from \cite{T3}\cite{PT}.

\section{Correlation functions in Liouville theory and the 
$H_3^+$ WZNW model}

\subsection{Five point function in Liouville theory}

As indicated above, we will heavily use properties of certain 
correlation functions in Liouville theory. 
Let $V_{\al}(z)$ denote local primary fields in Liouville theory
with conformal dimension $\De_\al=\al(Q-\al)$, $Q=b+b^{-1}$, 
where $b$ is the basic parameter (``coupling constant'') 
that determines the central charge of the 
Virasoro algebra according to $c=1+6Q^2$. 

We shall consider the vaccuum expectation value 
\begin{equation}
\Omega_{\rm Liou}^{(4)}\bigl(
\begin{smallmatrix} \al_3 & \al_2 \\
\al_4 & \al_1 \end{smallmatrix} 
|x,z\bigr)\;\equiv\;
\bra V_{\al_4}(\infty)V_{\al_3}(1)V_{-\frac{1}{2b}}(x)V_{\al_2}(z)
V_{\al_1}(0)\ket_{\rm Liou}^{}.
\end{equation}
\newcommand{\mal}{\begin{smallmatrix} \al_3 & \al_2 \\
\al_4 & \al_1 \end{smallmatrix}} 
It can be represented as follows
\cite{ZZ}\cite{T3}:
\begin{equation}\label{fourptLIOU}\begin{aligned}
\Omega_{\rm Liou}^{(4)}\bigl(
\begin{smallmatrix} \al_3 & \al_2 \\
\al_4 & \al_1 \end{smallmatrix} 
|x,z\bigr)\;=\; &\sum_{s=+,-}\int\limits_{0}^{\infty}dP\;
C(\al_4,\al_3,\al_P\!-\!\fr{s}{2b})C_s(\al_P)C(\al_{-P},\al_2,\al_1)\ti\\
 & \hspace{4cm}\ti\CF_{\al_P}^s\bigl(\mal|x,z\bigr)
\CF_{\al_P}^s\bigl(\mal|\bx,\bz\bigr),
\end{aligned}\end{equation}
where $\al_P=\frac{Q}{2}+iP$.
This expression is composed out of the following ingredients:
\begin{itemize}
\item The three point function $C(\al_3,\al_2,\al_1)$ \cite{DO}\cite{ZZ}:
\begin{equation}\label{ZZform}\begin{aligned}
C( & \alpha_1,  \alpha_2,\alpha_3)=\left[\pi\mu\gamma(b^2)b^{2-2b^2}
\right]^{(Q-\sum_{i=1}^3\alpha_i)/b}\times\\
& \times \frac{\Upsilon_0\Upsilon(2\alpha_1)\Upsilon(2\alpha_2)
\Upsilon(2\alpha_3)}{
\Upsilon(\alpha_1+\alpha_2+\alpha_3-Q)
\Upsilon(\alpha_1+\alpha_2-\alpha_3)\Upsilon(\alpha_2+\alpha_3-\alpha_1)
\Upsilon(\alpha_3+\alpha_1-\alpha_2)}.
\end{aligned}\end{equation}
The special function $\up(x)$ that appears in \rf{ZZform}
can be defined by means of the integral representation
\begin{equation}\label{upint}
\log\Upsilon(x)=
\int\limits_0^\infty\frac{dt}{t}\;\biggl[\biggl(\frac{Q}{2}-x\biggr)^2e^{-t}
-\frac{\sinh^2\bigl(\frac{Q}{2}-x\bigr)\frac{t}{2}}{\sinh\frac{bt}{2}
\sinh\frac{t}{2b}}\biggr].
\end{equation}
\item The structure constants $C_s(\al)$, $s=+,-$ are defined 
from the short-distance asymptotics
\begin{equation}
V_{-\frac{1}{2b}}(x)V_{\al}(z)\;\underset{x\ra z}{\sim}\;
\sum_{s=+,-}|x-z|^{2(\De_{\al_s}-\de-\De_{\al})}
V_{\al-\frac{s}{2b}}(z),
\end{equation}
where $\al_s\equiv \al-\frac{s}{2b}$, $\de\equiv\De_{\frac{1}{2b}}$.
The $C_s(\al)$, $s=+,-$
can be recovered as particular residues from the three point function
$C(\al_3,\al_2,\al_1)$ \cite{ZZ}\cite{T3}:
\begin{equation}
C_+(\al)\;=\;1,\qquad 
C_-(\al)\;=\; \bigl[\pi \mu \ga(b^2)b^{2-2b^2}\bigr]^{b^{-2}}b^{-2-2b^{-2}}
\frac{\ga(2b^{-1}\al-1-b^{-2})}{\ga(2b^{-1}\al)},
\end{equation}
where $\ga(x)=\Ga(x)/\Ga(1-x)$. 
\item The conformal blocks $\CF_{\al}^s\bigl(\mal|x,z\bigr)$, $s=+,-$:
They can be characterized as those solutions to the null vector decoupling 
equations 
\begin{equation}\label{decoup}\begin{aligned}
\biggl(b^2\pa_x^2+\frac{z(z-1)}{x(x-1)(x-z)}& \pa_z+
\frac{1-2x}{x(x-1)}\pa_x-\\
 -& \frac{\De}{x(x-1)}+\frac{\De_{\al_3}}{(x-1)^2}
+\frac{\De_{\al_2}}{(x-z)^2}+\frac{\De_{\al_1}}{x^2}\biggr)\CF(x,z)=0,
\end{aligned}\end{equation}
$\De=\De_{\al_1}+\De_{\al_2}+\De_{\al_3}+\de-\De_{\al_4}$, 
which have the following asymptotic behavior for $z\ra 0$:
\begin{equation}
\CF_{\al}^s\bigl(\mal|x,z\bigr)\;\underset{z\ra 0}{\sim}
\; z^{\De_{\al}-\De_{\al_2}-\De_{\al_2}}\CF_{\al}^s\bigl(\smal|x\bigr),
\end{equation}
with $\CF_{\al}^s\bigl(\smal|x\bigr)$ given by
\begin{equation}\begin{aligned}
\CF_{\al}^+\bigl(\smal|x\bigr)=&
z^{b^{-1}\al}(1-z)^{b^{-1}\al_3}F(u,v,w;z) \\
\CF_{\al}^-\bigl(\smal|x\bigr)=&z^{b^{-1}(Q-\al)}
(1-z)^{b^{-1}\al_3}F(u-w+1,v-w+1,2-w;z),
\end{aligned}\end{equation}
where we have used the notation
\[  \begin{aligned}
u=& b^{-1}(\al+\al_3+\al_4-3b^{-1}/2)-1\\
v= & b^{-1}(\al+\al_3-\al_4-b^{-1}/2)
\end{aligned}\qquad
w=b^{-1}(2\al-b^{-1}).
 \] 

\end{itemize}

For Liouville theory it is possible to prove
crossing symmetry and locality \cite{T3}.
This implies that $\Omega_{\rm Liou}\bigl(
\begin{smallmatrix} \al_3 & \al_2 \\
\al_4 & \al_1 \end{smallmatrix} 
|x,z\bigr)$ is real analytic on $\{(x,z)\in\BP_*^1\ti\BP_*^1;
z\neq x\}$, where $\BP_*^1\equiv \BP\setminus\{0,1,\infty\}$, and that
the following identity holds:
\begin{equation}\label{LIOUcross}
\Omega_{\rm Liou}^{(4)}\bigl(
\begin{smallmatrix} \al_3 & \al_2 \\
\al_4 & \al_1 \end{smallmatrix} 
|x,z\bigr)\;=\;
\Omega_{\rm Liou}^{(4)}\bigl(
\begin{smallmatrix} \al_1 & \al_2 \\
\al_4 & \al_3 \end{smallmatrix} 
|1-x,1-z\bigr).
\end{equation}

\begin{rem}
We do not have an explicit expression for the functions 
$\CF_{\al}^s\bigl(\mal|x,z\bigr)$. However, the free field construction of 
chiral vertex operators intertwining between three general irreducible
highest weight representations of the Virasoro algebra given in \cite{T3}
leads to a constructive definition for $\CF_{\al}^s\bigl(\mal|x,z\bigr)$
that is manageable enough to allow for the calculation of the 
monodromies of $\CF_{\al}^s\bigl(\mal|x,z\bigr)$. These turn out to
be related to canonical objects from quantum group representation theory, 
the so-called b-Racah-Wigner coefficients \cite{PT}. The identities 
needed to prove \rf{LIOUcross} all follow from the definition of the 
b-Racah-Wigner coefficients in terms of quantum group representation theory
\cite{PT}\cite{T3}.
\end{rem}

\subsection{Four point function in the $H_3^+$ WZNW model}

The basic parameter in the $H_3^+$ WZNW model is the level $k$ of the
current algebra 
\begin{equation}\label{algebra}
\begin{aligned} {[}J_n^0,J_m^0{]}& =-\fr{k}{2}n\de_{n+m,0} \\ 
{[}J_n^0,J_m^{\pm}{]}& =\pm J_{n+m}^{\pm} \end{aligned} \qquad
{[}J_n^{-},J_m^{+}{]}=2J_{n+m}^0+kn\de_{n+m,0}.  
\end{equation}
It will be useful to write $k$ as $k\equiv b^{-2}+2$, since $b$ will
turn out to correspond directly to the parameter $b$ of Liouville 
theory. The definition of primary fields $\Phi^j(x|z)$ will be the one used
in \cite{T2}, the conformal dimension being $\De_j=-b^2j(j+1)$.

We will be interested in the following
correlation function in the $H_3^+$ WZNW model:
\newcommand{\mj}{\begin{smallmatrix} j_3 & j_2 \\
j_4 & j_1 \end{smallmatrix}} 
\begin{equation}
\Omega_{\sst\rm WZNW}^{(4)}\bigl(\mj|x|z\bigr)\;\equiv \;
\bra \Phi^{j_4} (\infty|\infty)\Phi^{j_3}(1|1)  \Phi^{j_2}(x|z) 
\Phi^{j_1}(0|0)\ket^{}_{\rm\sst WZNW}
\end{equation}
The following description was proposed in \cite{T2} for 
$\Omega_{\sst\rm WZNW}^{(4)}\bigl(\mj|x|z\bigr) $: 
\begin{equation}\label{fourptWZNW}\begin{aligned}
\Omega_{\sst\rm WZNW}^{(4)}\bigl(\mj|x|z\bigr)\; 
=\;\int\limits_{0}^{\infty}
\frac{dp}{B(j_p)}D(j_4,j_3,j_p)D(j_p,j_2,j_1)H_{j_p}\bigl(\mj|x|z\bigr).
\end{aligned}
\end{equation}
where $j_p=-\frac{1}{2}+ip$.
The following objects appear in \rf{fourptWZNW}:
\begin{itemize}
\item
The two-point function $B(j)$ is given by the formula
\begin{equation}\label{Bcoeff}
 B(j)= -\bigl(\nu(b)\bigr)^{2j+1}\frac{2j+1}{\pi}
\frac{\Ga(1+b^2(2j+1)\bigr)}{\Ga(1-b^2(2j+1)\bigr)}.
\end{equation}
The expression for $\nu(b)$ can be found in \cite{T2}, but will not be 
important in the following.
\item The three point function $D( j_3,j_2,j_1)$ has the following
expression:
\begin{equation}\label{Strconst}\begin{aligned}
D&(j_3,j_2,j_1) \;=\\
& \!=\; \frac
{(\nu(b)b^{2b^2})^{j_1+j_2+j_3+1}\; C_{\sst\rm W}(b) \;
\up_{\sst\rm W}(2j_1+1)\up_{\sst\rm W}(2j_2+1)\up_{\sst\rm W}(2j_3+1)}
{\up_{\sst\rm W}(j_1+j_2+j_3+1)\up_{\sst\rm W}(j_1+j_2-j_3)\up_{\sst\rm W}(j_1+j_3-j_2)
\up_{\sst\rm W}(j_2+j_3-j_1)}.
\end{aligned}\end{equation}
The special function $\up_{\sst\rm W}(j)$ is related to the $\Upsilon$-function
via $\up_{\sst\rm W}(j)\equiv\Upsilon(-bj)$.
\item The function $H_j\bigl(\mj|x|z\bigr)$ can be decomposed into
{\it conformal blocks} $\CG_{j}\bigl(\mj|x|z\bigr)$
as follows:
\begin{equation}\label{hjdecomp}\begin{aligned}
H_j\bigl(\mj|x|z\bigr)\;=& \;\CG_{j}\bigl(\mj|x|z\bigr)
\CG_{j}\bigl(\mj|\bx|\bz\bigr)-\\
 - & \frac{(2j+1)^2}{\ga^2(2j+2)}
\frac{\ga(j_4-j_3+j+1)\ga(j_3-j_4+j+1)}{\ga(j_2-j_1-j)\ga(j_1-j_2-j)}\ti\\
& \hspace{3cm}\ti
\CG_{-j-1}\bigl(\mj|x|z\bigr)\CG_{-j-1}\bigl(\mj|\bx|\bz\bigr)
\end{aligned}\end{equation}
\item The conformal blocks $\CG_{j}\bigl(\mj|x|z\bigr)$ 
are uniquely defined as 
those solutions of the Knizhnik-Zamolodchikov- (KZ-) equations 
$\bigl(z(z-1)\pa_z+ b^2\CD_x^{(2)}\bigr)\CG=0$,  
\begin{equation}\label{kzred}\begin{split}
\CD_x^{(2)} =\;\; & x(x-1)(x-z)\pa_x^2 \\
            - & \bigl[(\kappa-1)(x^2-2zx+z)+2j_1x(z-1)+2j_2x(x-1)+
                2j_3z(x-1)\bigr]\pa_x \\
            + &2j_2\kappa(x-z)+2j_1j_2(z-1)+2j_2j_3z,
\end{split}\end{equation}
which can be represented by power series of the form
\begin{equation}\label{pwser} \CG_{j}\bigl(\mj|x|z\bigr)=
z^{\De_{21}(j)}\;\sum_{n=0}^{\infty}\; z^n\;
\CG_{j}^{(n)}\bigl(\mj|x\bigr),
\end{equation}
with initial term 
$G_{j}^{(0)}\bigl(\mj|x\bigr)$ given by
\begin{equation}\label{zerodef}
G_{j}^{(0)}\bigl(\mj|x\bigr)
\equiv x^{j_1+j_2-j}F(j_1-j_2-j,j_4-j_3-j;-2j;x).
\end{equation}
We have used the abbreviations $\kappa=j_1+j_2+j_3-j_4$ and
$\De_{21}(j)=\De_{j}-\De_{j_2}-\De_{j_1}$.
\end{itemize}
Our aim will be to prove that
that $\Omega_{\sst\rm WZNW}^{(4)}\bigl(
\begin{smallmatrix} \al_3 & \al_2 \\
\al_4 & \al_1 \end{smallmatrix} 
|x|z\bigr)$ is real analytic on $\{(x,z)\in\BP_*^1\ti\BP_*^1;
z\neq x\}$, and that
the following identity ({\it crossing symmetry}) holds:
\begin{equation}
\Omega_{\rm\sst WZNW}^{(4)}\bigl(\mj
|x|z\bigr)\;=\;
\Omega_{\rm\sst WZNW}^{(4)}\bigl(
\begin{smallmatrix}j_1 & j_2 \\
j_4 & j_3 \end{smallmatrix} 
|1-x|1-z\bigr).
\end{equation}

\section{Proof of crossing symmetry}

Crossing symmetry for the four point function in the 
$H_3^+$ WZNW model will follow 
immediately once the following identity is proven:
\begin{equation}\label{Liou-WZNW}
\begin{aligned}
\Omega_{\sst\rm WZNW}^{(4)}\bigl(\mj|x|z\bigr)\;=& \;
\Omega_{\sst\rm AUX}^{(4)}\bigl(\mj|x,z\bigr)
\Omega_{\sst\rm LIOU}^{(4)}\bigl(\mal|x,z\bigr),\\
\Omega_{\sst\rm AUX}^{(4)}\bigl(\mj|x,z\bigr)\;\equiv\;
& \;\frac{\pi}{b^2}\frac{C_{\sst\rm W}^2(b)}{\up_0^2}\frac
{(\nu(b)b^{2b^2})^{t}}{(\pi\mu\gamma(b^2)b^{2-2b^2}
)^{s}_{}}
\prod_{i=1}^4\frac{\up_{\sst\rm W}(2j_i+1)}{\up(2\al_i)}\ti\\
&\ti |x|^{-2b^{-1}\al_1}|1-x|^{-2b^{-1}\al_3}|x-z|^{-2b^{-1}\al_2}
|z|^{-\ga_{12}}|1-z|^{-\ga_{23}},
\end{aligned}\end{equation}
where we have used the notation
\begin{equation}\label{defs}\begin{aligned}
s=& b^{-1}\Bigl(Q-\sum_{i=1}^4\alpha_i+\frac{1}{2b}\Bigr),\\
\ga_{23} = & 4b^2j_2j_3-4\al_2\al_3,
\end{aligned}\qquad\begin{aligned}
t=& \sum_{i=1}^4j_i+1,\\
\ga_{12} = & 4b^2j_1j_2-4\al_1\al_2,
\end{aligned}\end{equation} 
and assumed the following identifications between the
sets of variables $j_4,\dots,j_1$ and $
\al_4,\dots,\al_1$:
\begin{equation}\label{varident}\begin{aligned}
2\al_1=& -b(j_1+j_2-j_3-j_4-b^{-2}-1),\\
2\al_2=& -b(j_1+j_2+j_3+j_4+1),
\end{aligned}\qquad\begin{aligned}
2\al_3=& -b(j_2+j_3-j_1-j_4-b^{-2}-1),\\
2\al_4=& -b(j_4+j_2-j_1-j_3-b^{-2}-1).\\
\end{aligned}\end{equation}
Crossing symmetry of $\Omega_{\sst\rm WZNW}^{(4)}\bigl(\mj|x|z\bigr)$
then follows from the corresponding properties of
$\Omega_{\sst\rm LIOU}^{(4)}\bigl(\mal|x|z\bigr)$ and
$\Omega_{\sst\rm AUX}^{(4)}\bigl(\mj|x|z\bigr)$.

The main ingredient needed to prove identitity \rf{Liou-WZNW}
is the following observation from \cite{FZ}: 
\begin{itemize}
\item[] {\it Assume that (i) $j_4,\dots,j_1$ and $
\al_4,\dots,\al_1$ satisfy the relations \rf{varident}, 
(ii) $\ga_{23}$ and $\ga_{12}$ are defined by \rf{defs}, and 
(iii) $\CF(x,z)$ and $\CG(x|z)$ are related by 
\begin{equation}\label{cfbl-ident} \CF(x,z)\;=\;
x^{b^{-1}\al_1}(1-x)^{b^{-1}\al_3}(x-z)^{b^{-1}\al_2}
z^{\frac{1}{2}\ga_{12}}(1-z)^{\frac{1}{2}\ga_{23}}\CG(x|z).
\end{equation}
$\CF(x,z)$ will then satisfy the decoupling equations \rf{decoup} 
if and only if $\CG(x|z)$ solves the KZ-equations 
$\bigl(z(z-1)\pa_z \CF - b^2\CD_x^{(2)}\bigr)\CF=0$.}
\end{itemize}
This observation reduces the proof of \rf{Liou-WZNW} to 
a couple of straightforward verifications. First let us note that 
\begin{equation}\label{id1}
G_{j}^{(0)}\bigl(\mj|x\bigr)=
x^{b^{-1}(\al_1+\al_2)}(1-x)^{b^{-1}\al_3}
\CF_{\al}^s\bigl(\smal|x\bigr)
\end{equation}
if the identifications \rf{varident} hold, and that 
\begin{equation}\label{id2}
z^{\De_{\al}-\De_{\al_2}-\De_{\al_2}}\;=\;
z^{\De_{j}-\De_{j_2}-\De_{j_2}}z^{\frac{1}{2}\ga_{12}}
\end{equation}
if $\ga_{12}$ is chosen according to \rf{defs} and the relation
between $j$ and $\al$  is assumed to be
\begin{equation}\label{al-jident}
\al\;=\;-bj+\frac{1}{2b}.
\end{equation}
But this implies that 
\begin{equation}\label{cfbl-ident2} \CF_{\al}^s\bigl(\mal|x,z\bigr)\;=\;
x^{b^{-1}\al_1}(1-x)^{b^{-1}\al_3}(x-z)^{b^{-1}\al_2}
z^{\frac{1}{2}\ga_{12}}(1-z)^{\frac{1}{2}\ga_{23}}
\CG_{j}\bigl(\mj|x|z\bigr).
\end{equation}
Indeed, eqns. \rf{id1} and \rf{id2} imply that both sides of 
\rf{cfbl-ident2} have the same asymptotic behavior for 
$z\ra 0$. Power series solutions to both the decoupling 
equation \rf{decoup} and the KZ-equations 
are uniquely determined by their initial terms 
$\CF_{\al}^s\bigl(\smal|x\bigr)$
and $G_{j}^{(0)}\bigl(\mj|x\bigr)$ respectively
\footnote{See \cite[Appendix C]{T2} for a verification of this statement
in the case of the KZ-equation which can easily be adapted to the 
decoupling equation \rf{decoup}.}. 
\rf{id1} and \rf{id2} are therefore enough to conclude that
\rf{cfbl-ident2} indeed holds.

In order to prove \rf{Liou-WZNW}, it therefore suffices
to consider the coefficients with which the conformal blocks are 
multiplied in \rf{fourptLIOU} and \rf{fourptWZNW}.
Let us first note that 
\begin{equation}\label{strconstid}
\frac{C(\al_4,\al_3,\al\!-\!\fr{1}{2b})C(\al,\al_2,\al_1)}
{D(j_4,j_3,j)B^{-1}(j)D(j,j_2,j_1)}\;=\;
\frac{b}{\pi}\frac{\up_0^2}{C_{\sst\rm W}^2(b)}\frac
{(\pi\mu\gamma(b^2)b^{2-2b^2}
)^{s}_{}}{(\nu(b)b^{2b^2})^{t}}
\prod_{i=1}^4\frac{\up(2\al_i)}{\up_{\sst\rm W}(2j_i+1)}.
\end{equation}
The verification of \rf{strconstid} is done by a straightforward calculation
using 
the explicit expressions for $B$, $C$ and  $D$ given above, the identifications
\rf{varident}, \rf{al-jident} and the functional equations \cite{ZZ}
\begin{equation}\label{funids}
\up(x+b)=b^{1-2bx}\ga(bx)\up(x),\qquad
\up(x+b^{-1})=b^{-1+2b^{-1}x}\ga(b^{-1}x)\up(x).
\end{equation}
The final point that remains to prove \rf{Liou-WZNW}
is then to verify the fact that 
\begin{equation}\label{strconstid2}
\frac{C(\al_4,\al_3,\al\!+\!\fr{1}{2b})C_-(\al)}
{C(\al_4,\al_3,\al\!-\!\fr{1}{2b})C_+(\al)}\;=\;-
\frac{(2j+1)^2}{\ga^2(2j+2)}
\frac{\ga(j_4-j_3+j+1)\ga(j_3-j_4+j+1)}{\ga(j_2-j_1-j)\ga(j_1-j_2-j)},
\end{equation}
cf. \rf{hjdecomp}, which is again a matter of straightforward 
computations using \rf{funids}.

\end{document}